\documentclass[jcpbfk,article,traditional]{achemso}

\usepackage{epsfig}
\usepackage{amsfonts}
\usepackage{color}
\usepackage{amsmath}

\title{Dynamic Heterogeneity In The Monoclinic Phase Of CCl$_4$}

\author{Nirvana B. Caballero }
\email{ncaballe@famaf.unc.edu.ar  }
\affiliation{Facultad de Matem\'atica, Astronom\'{\i}a y F\'{\i}sica,
Universidad Nacional de C\'ordoba, C\'ordoba, Argentina and IFEG-CONICET,
Ciudad Universitaria,
X5016LAE C\'ordoba, Argentina}

\author{Mariano Zuriaga }
\email{zuriaga@famaf.unc.edu.ar}
\affiliation{Facultad de Matem\'atica, Astronom\'{\i}a y F\'{\i}sica,
Universidad Nacional de C\'ordoba, C\'ordoba, Argentina and IFEG-CONICET,
Ciudad Universitaria,
X5016LAE C\'ordoba, Argentina}

\author{Marcelo Carignano}
\email{mcarignano@qf.org.qa}
\affiliation{Qatar Environment and Energy Research Institute, 
Hamad Bin Khalifa University, Qatar Foundation, P.O. Box 5825, Doha, Qatar }

\author{Pablo Serra}
\email{serra@famaf.unc.edu.ar}
\affiliation{Facultad de Matem\'atica, Astronom\'{\i}a y F\'{\i}sica,
Universidad Nacional de C\'ordoba, C\'ordoba, Argentina and IFEG-CONICET,
Ciudad Universitaria,
X5016LAE C\'ordoba, Argentina}
\phone{+54(0)3514334051, ext. 218}
\fax{+54(0)3514334054}

\begin{document}

\newpage

\begin{abstract}
Carbon tetrachloride (CCl$_4$) is one of the simplest
compounds having a translationally stable monoclinic phase while
exhibiting a rich rotational dynamics below 226 K. Recent nuclear
quadrupolar resonance (NQR) experiments revealed that the dynamics of CCl$_4$
is similar to that of the other members of the isostructural
series CBr$_{n}$Cl$_{4-n}$, suggesting that
the universal relaxation features of canonical glasses such as
$\alpha$- and $\beta$-relaxation are also present in non-glass
formers. Using molecular dynamics (MD) simulations we studied the
rotational dynamics in the monoclinic phase of CCl$_4$.
The molecules undergo $C3$ type jump-like rotations around each one
of the four C-Cl bonds. The rotational dynamics is very well described
with a master equation using as the only input the rotational rates
measured from the simulated trajectories. It is found that the
heterogeneous dynamics emerges from faster and slower modes
associated with different rotational axes, which have fixed
orientations relative to the crystal and are distributed
among the four non-equivalent molecules of the unit cell.
\end{abstract}
\date{\today}

\newpage

\section{Introduction}

The currently accepted scenario for canonical glasses includes
different relaxation mechanisms that are universally present in all
systems. Experimentally, these different mechanisms are clearly revealed
by the dielectric spectra that shows a broad low frequency peak
referred to as $\alpha$-relaxation
\cite{Lunkenheimer_2000,Lunkenheimer_2002} and a higher frequency peak
or shoulder usually called Johari-Goldstein $\beta$-relaxation
\cite{Affouard_2005,Ngai_2004,Pardo_2006,Johari_1970,Ngai_1998,Schneider_2000,Johari_2002,Johari_1973,Capaccioli_2007,Capaccioli_2011,
JimenezRuiz_PRL_1999,JimenezRuiz_PRB_1999}.
The $\alpha$-relaxation is generally attributed to processes involving
the cooperative dynamics of regions of molecules 
\cite{Lunkenheimer_2002,Lunkenheimer_2000}.
The microscopical origin of the $\beta$ relaxation is still a matter
of wide debate\cite{Ngai_2004,Pardo_2006,Johari_1970,Ngai_1998,Schneider_2000,Johari_2002,Johari_1973,Capaccioli_2007,Capaccioli_2011,JimenezRuiz_PRL_1999}.
The proposed models explain this peak as a consequence of the nonuniformity
of the glassy state involving only local regions in
which molecules can diffuse (islands of mobility).
An alternative homogeneous explanation attributes the
secondary relaxation phenomena to small-angle reorientations of all the
molecules \cite{Romanini_2012,Voguel_2000,Voguel_2001}.

Systems having translational crystalline order but rotational degrees
of freedom also display a glassy behavior. Since in these cases one
type of degree of freedom is completely absent, they represent
simplified models on which to test theoretical concepts on glassy dynamics
\cite{Drozd-Rzoska_PRB_2006,MartinezGarcia_JCP_2011,MG_JCP_2011,MartinezGarcia_JCP_2010}.
Compounds of molecules of the type
CBr$_{n}$Cl$_{4-n}$, with n=0,1,2, are
examples of systems with this characteristics
\cite{Zuriaga_PRL_2009,Zuriaga_JCP_2012,Caballero_2012}.
They have a series of
solid-solid phase transitions attributed to the ability of the
molecules to acquire rotational degrees of freedom as the temperature
is increased. All these compounds crystallize from the melting to a
plastic FCC phase. Further reduction of the temperature leads to a 
$C2/c$ monoclinic phase
\cite{Pardo_JCP_1999,Pardo_PCCP_2001,Pardo_JPCB_2001,Pardo_JCP_2000,Pothoczki_PRB_2012}.
The cases n=1 and n=2 exhibit a glass transition at
90 K that is clearly visible through calorimetric techniques 
and its structure shows disorder
in the position of the Cl and Br atoms \cite{Ohta_JPC_1995,Binbrek_MolPhys_1999}.
The case n=0 cannot
display such a disorder and, in fact, the calorimetric curve does not
show a glass transition at low temperatures \cite{Ohta_JPC_1995}.

The dielectric spectra of CBrCl$_{3}$ and
CBr$_{2}$Cl$_{2}$ was reported by Zuriaga {\em et al.}
\cite{Zuriaga_PRL_2009} in the temperature range 100 -- 250 K and
100 -- 210 K, respectively. At the lower end of the temperature range
the most relevant characteristic is that both spectra display a
well-defined shoulder on the high frequencies flank of the 
$\alpha$-peak, which is attributed to the $\beta$-relaxation.
Since CCl$_4$ has no molecular dipole moment it is not
accessible to dielectric experiments but it can be studied using
nuclear quadrupolar resonance.
Interestingly, the resolution of the
NQR spectra for CCl$_4$ is well superior to the
corresponding spectra for CBr$_{2}$Cl$_{2}$ and
CBrCl$_{3}$ and the two techniques complement each other.
On the other hand, the NQR experiments are limited to a temperature
range between 77 K and 140 K, with the upper end determined by the
broadening of the signal. The picture that emerges from the combined
analysis is that the three compounds have a very similar dynamic
evolution in the monoclinic phase as a function of temperature
\cite{Zuriaga_PRL_2009,Zuriaga_JCP_2012}.
The analysis of the isostructural CCl$_4$ shows that nonequivalent
molecules in the unit cell perform reorientational
jumps at different time scales due to their different
crystalline environments.
These results support the conclusion that the dynamic heterogeneity
is intimately related to the secondary relaxation observed
in these compounds \cite{Zuriaga_PRL_2009,Zuriaga_JCP_2011,Zuriaga_JCP_2012}.

This work is a direct extension of our previous works involving halogenomethanes  \cite{Zuriaga_PRL_2009,Zuriaga_JCP_2011,Zuriaga_JCP_2012}. In this case we have studied the monoclinic phase of CCl$_4$ using extensive molecular dynamics simulations on a fully anisotropic cell, and the results were compared with an analytical stochastic model. The advantage of this particular system is that it allows for the determination of rotational correlation times from NQR measurements with sufficient precision to distinguish between the nonequivalent groups of molecules. The same distinction is easily done in the molecular dynamics simulations, which in turn allows for an analysis at the individual nonequivalent groups and correlate the results with the location of the molecules in the lattice. The temperature range affordable by the simulations covers from the FCC phase down to 160 K. Simulations at lower temperatures became impossible as the relaxation times go beyond 10 $\mu$s. Is is found that, when the spin-lattice relaxation times from the experiments and from the simulations are plotted in the 100 -- 220 K temperature range, the two sets of curves correspond to each other very well for the five distinguishable modes suggesting that CCl$_4$ behaves as a strong glass in the whole temperature range, in spite of the fact that the compound cannot glassify due to its lack of orientational entropy. 

\section{ Theoretical Methods }
\subsection*{Model and computational details}

CCl$_4$ is a tetrahedral molecule having three equivalent
$C_2$ and four equivalent $C_3$ symmetry axes. Following our previous
work\cite{Zuriaga_JCP_2011}, we have modeled the CCl$_4$
molecule as a rigid, non-polarizable tetrahedron with the carbon atom
at its center and a chlorine atom at each one of the vertices. The
interaction between molecules is represented by a combination of
Lennard-Jones and Coulombic terms summarized in Table
\ref{table:potential}. The cross interaction between atoms of
different type is calculated by applying the Lorentz-Berthelot
combination rules, i.e., geometrical mean for $\epsilon$ and
arithmetic mean for $\sigma$. A spherical cut-off at 1.5 nm was
imposed on all intermolecular interactions. Periodic boundary
conditions were imposed in all three Cartesian directions.

\begin{table}[ht]
\caption{\label{table:potential} CCl$_4$ model parameters and geometry.}                        
\centering  
\begin{tabular}{l c c c c c}
\hline\hline \\[-2.0ex]                                                                                                                                      
  & $\epsilon$ [kJ/mol] & $\sigma$ [nm] & $q$ [e] & \multicolumn{2}{c}{Bond [nm]} \\ 
\hline
C & 0.22761 & 0.37739 & -0.696 & C-Cl & 0.1766 \\
Cl& 1.09453 & 0.34667 & 0.174  & Cl-Cl& 0.2884 \\
\hline\hline
\end{tabular}
\end{table}

The monoclinic crystal structure of CCl$_4$, resolved by
Cohen et {\em al.} at 195 K \cite{Cohen_ActaCryst_1979}, corresponds
to the $C2/c$ space group. The unit cell, which contains $Z$=32
molecules, has the following lattice parameters: $a$=2.0181 nm,
$b$=1.1350 nm, $c$=1.9761 nm and angle $\beta=111.46^{\circ}$.
Using the experimental crystalline structure as initial coordinates,
we constructed a simulation supercell containing  512 molecules, which
correspond to 16 monoclinic unit cells. This supercell was prepared by
replicating the experimental unit cell twice on the $x$ and $z$
directions and four times in the $y$ direction.

The molecular dynamics simulations, conducted under NPT conditions,
have been carried out using the Gromacs v5.0.2 simulation package.
Atom-atom distances within each molecule were kept constants with the
SHAKE algorithm. The simulations were started with a 10 ns
equilibration run with the pressure controlled by a Berendsen barostat
and the temperature controlled with the v-rescale thermostat. The
Berendsen weak coupling method ensured a smooth approach to
equilibrium with no disruptions to the simulation cell. The classical
Newton's equations were integrated using the leap-frog algorithm and
the time step of the integration of the equations of motion was set
to 1 fs. 

The production runs were extended up to 100 ns or 10 $\mu$s depending
on the temperature, using in this case a time step of 5 fs. The
control of the temperature was made by using a Nos\'e-Hoover
thermostat, with a time constant of 2.0 ps.
The pressure was maintained constant by using a fully anisotropic
Parrinello-Rahman barostat with a reference pressure of 1 atm.
The study covered temperatures ranging from 160 K to 230 K, in steps of 10 K.
The monoclinic structure was stable in the whole
analyzed temperature range (see Supporting Information file).

\section{Results and Discussion}

\begin{figure}[t]
\begin{center}
\includegraphics[width=0.4\textwidth]{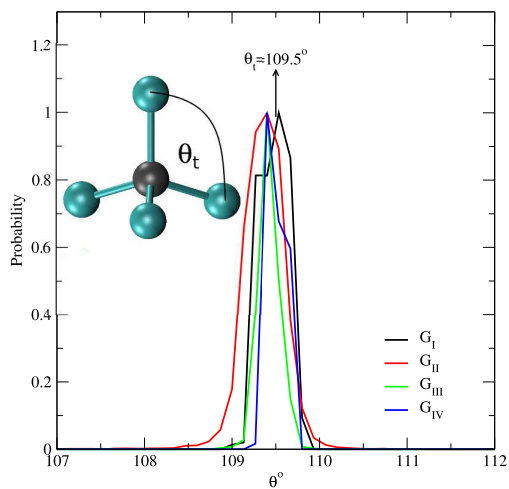}
\end{center}
\caption{\label{angles_160} 
Normalized frequency distribution of reorientation angles for the
four distinct group of molecules, as indicated. The results
correspond to the production run at 160 K. The maximum at the
tetrahedral angle indicate that all jumps are $C3$ type rotations.}
\end{figure}

The spatial positions of the molecules in the unit cell of the system
are defined through the application of 8 symmetry operations over the
4 nonequivalent molecules (16 Cl atoms with nonequivalent positions).
As a consequence, the system has four distinctive groups of molecules
that we will refer to as G$_{I}$, G$_{II}$,
G$_{III}$ and G$_{IV}$. In our simulation cell,
each group contains 128 molecules. Each molecule belonging to a given
group has the same specific arrangement of neighboring molecules. All
the molecular reorientation processes occurring during the simulations
are sudden large-angle jumps of the Cl atoms. The jumps were detected
using a running test algorithm based on a
signal to noise measure and
adapted from the work of Carter and Cross\cite{Carter_2005}. In this
method, the molecular jumps are detected by a spike in the test
function as explained in the Supporting Information file. After
analyzing the trajectories for all the temperatures, the time $t_i$ of
every single jump was registered. The angle described by a C-Cl bond
upon a jump was calculated using the average direction of the bond
before and after the jump. Namely, defining $\vec{b}_j$ as the bond
between the central C atom and the $j$-th Cl atom of the same
molecule, the angle jump at time $t_i$ is defined as the
angle between 
$\langle \vec{b}_j \rangle_{i-1}$ and $\langle \vec{b}_j \rangle_{i}$.
Here, the angular brackets represent time average and the subscript $i$
indicates the lapse between $t_{i}$ and $t_{i+1}$. In Figure
\ref{angles_160} we show the relative frequency of the reorientation
angles calculated from the simulation at 160 K and for the four
different groups of molecules. All the curves peak at the tetrahedral
angle, which is the angle that should be observed upon $C3$ type
rotations. Indeed, a careful inspection confirmed that for all
temperatures there were just a handful of cases corresponding to $C2$
type rotations and therefore there were neglected in the analysis.

In order to characterize the molecular rotations using standard self
correlation methods, we calculate the relaxation of the bond
orientation through self correlation functions defined as

\begin{equation}
C_j(t)=\langle \vec{b}_j(0).\vec{b}_j(t) \rangle,
\end{equation}
where $j=1,\dots,4$ represents a particular C-Cl bond,
$\vec{b}_j(0)$ represents the position of the C-Cl$_{j}$ bond
at the initial condition, and the angular brackets represent the average over
the 128 molecules of the same group.

\begin{figure}[t!]
\begin{center}
  \includegraphics[width=0.4\textwidth]{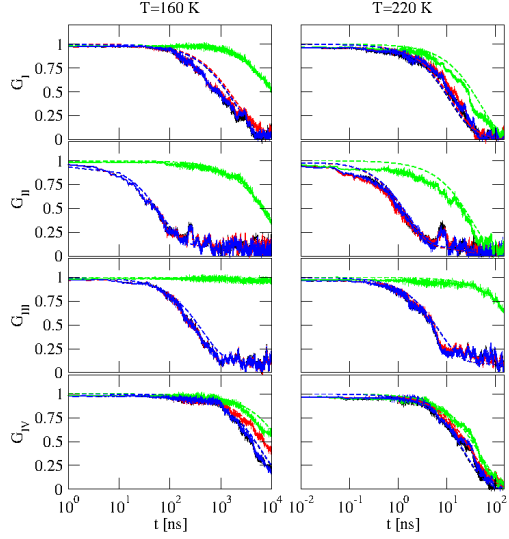}
\end{center}
\caption{ \label{cgroups}
Auto-correlation functions $C_j(t)$ of the 4 bonds C-Cl$_{j}$ 
averaged over molecules within the four groups of the system.
The solid lines correspond to the MD simulation results and the dashed
lines are derived from the master equation Eq. (\ref{master_equation}).
Different colors correspond to the four different bonds.}
\end{figure}

The resulting self correlation functions are shown in Figure
\ref{cgroups} with solid lines. The correlation times for the reorientations
of the molecules depend not only on the groups but also the curves
reveal that one bond (green lines) maintains its orientation for a
time considerably longer than the other three bonds.
This indicates that the molecules have preferential axes of rotation.
G$_{III}$ exhibit the strongest rotational anisotropy,
followed by G$_{II}$ and G$_{I}$ and only a
minor effect is observed in G$_{IV}$. The strength of this
anisotropic character decreases with increasing temperature as all
the free energy barriers become relatively smaller.

During a rotation of the type $C3$ one of the C-Cl bonds remain in
its place and therefore one Cl atom does not change its average
position. The remaining three Cl atoms undergo a jump of 109.5
degrees. Each molecule has four possible axis for a $C3$ type 
rotation that coincide with the C-Cl molecular bonds. As all the
molecular bonds remain lying around the four initial orientations in
the crystal it is possible to define, for each molecule, four axis of
rotation relative to the overall orientation of the crystal. In this
way we define the axes $a_i\,, i=1,\dots,4$, for each one of the
molecules in the simulation cell. The frequencies of rotational jumps
for the four groups of molecules along the four possible axes of
rotation, averaged over each group, are reproduced in Table
\ref{table:w}. The numerical values of the frequencies span over four
order of magnitude implying the existence of different dynamical
modes in the crystal as already suggested by the
rotational self correlation functions.

\begin{table*}[t!]
\begin{tiny}
\caption{\label{table:w} CCl$_4$ average reorientation 
frequencies $w_i$ ($i=1,\dots,4$) (in ns$^{-1}$) for the four groups of molecules about the
four possible directions of the axis of rotation.
For the fastest
rotational axes the inverse of the frequency $w_3$ is a good approximation
of the average waiting time between jumps ($\lambda^{-1}$).\\
}
\centering
\begin{tabular}{c|cccc|cccc}
\hline\hline \\[-2.0ex]                                                                                                                                      
\multicolumn{1}{c|}{} & \multicolumn{4}{c|}{\em 160 K} & \multicolumn{4}{c}{\em 220 K} \\ 
\hline
axis  & {G$_{I}$}& G$_{II}$ & G$_{III}$  & G$_{IV}$ & G$_{I}$ & G$_{II}$ & G$_{III}$ & G$_{IV}$ \\
$a_1$ & $3.91 \times10^{-6}$ & $7.81 \times10^{-7}$ & $1.56 \times10^{-6}$ & $7.81 \times10^{-7}$ & $2.80 \times10^{-3}$ & $1.62 \times10^{-4}$ & $3.23 \times10^{-4}$ & $1.51 \times10^{-3}$ \\
$a_2$ & $3.83 \times10^{-5}$ & $2.11 \times10^{-5}$ & $7.81 \times10^{-7}$ & $3.43 \times10^{-5}$ & $8.30 \times10^{-3}$ & $5.39 \times10^{-3}$ & $9.70 \times10^{-4}$ & $1.20 \times10^{-2}$ \\
$a_3$ & $3.62 \times10^{-4}$ & $1.00 \times10^{-2}$ & $1.52 \times10^{-3}$ & $7.42 \times10^{-5}$ & $4.76 \times10^{-2}$ & $4.26 \times10^{-1}$ & $9.36 \times10^{-2}$ & $1.78 \times10^{-2}$ \\
$a_4$ & $8.59 \times10^{-6}$ & $6.02 \times10^{-5}$ & $7.81 \times10^{-7}$ & $2.34 \times10^{-6}$ & $4.36 \times10^{-3}$ & $1.76 \times10^{-2}$ & $8.08 \times10^{-4}$ & $2.53 \times10^{-3}$ \\
\hline\hline
\end{tabular}
\end{tiny}
\end{table*}

In order to rationalize the behavior of the functions $C_j(t)$ we
developed a simple analytical model that derive these functions
from the relative reorientation frequencies. We propose a master
equation that describes the probabilities of occupation of the four
available sites in the molecule for the Cl atoms.
Let us define the
stochastic variables $Y_j$ that describe the position of the j-th Cl
atom on the four available sites. Then, each $Y_j$ takes values over a
range represented by  $\{1,2,3,4\}$, with a probability distribution
$p^j_n$ over this range. Assuming that a reorientational
jump in the molecule is a Markovian process, the master equation that describes
the probability $p^j_n$ of the state of $Y_j$ is

\begin{equation}
\label{master_equation}
\frac{dp^j_n(t)}{dt}= \sum_{n'} \left[ \nu_{nn'}p^j_{n'}(t)-\nu_{n'n}p^j_n(t) \right] \,; j=1,\dots,4,
\end{equation}
where $\nu_{nn'}$ is the transition probability per unit time from
the site $n'$ to the site $n$. These equations
may be written in matrix form as
\begin{equation}
\label{dp}
\frac{d\vec{P^j}}{dt}=V \vec{P^j},
\end{equation}
where $\vec{P^j}$ is the vector
$(p^j_1(t),p^j_2(t),p^j_3(t),p^j_4(t))$,
and $V$ is the $4 \times 4$ matrix
\begin{equation}
\left(
 \begin{array}{cccc}
    -(\nu_{21}+\nu_{31}+\nu_{41})  & \nu_{12} & \nu_{13} & \nu_{14}   \\
      \nu_{21}  & -(\nu_{12}+\nu_{32}+\nu_{42})  &\nu_{23} & \nu_{24} \\
      \nu_{31}  & \nu_{32} & -(\nu_{13}+\nu_{23}+\nu_{43}) & \nu_{34} \\
      \nu_{41}  & \nu_{42} & \nu_{43} & -(\nu_{14}+\nu_{24}+\nu_{34}) \\
\end{array}
\right).
\end{equation}

Eq. (\ref{dp}) is an homogeneous first order differential equation with constant coefficients.
The solution is
\begin{equation}
\vec{P^j}(t)=R^{-1} e^{V_D(t)}R\vec{P^j}(0)
\end{equation}
where $R$ is the matrix that diagonalize $V$, $V_D=RVR^{-1}$.
Assuming that C3
jumps are the exclusive mechanism of molecular reorientation,
the $\nu_{nn'}$ can be approximated using the frequency of rotations
$w_i$
around the axes $a_i$. These frequencies were calculated from the MD
simulations and are summarized
in Table \ref{table:w}.
They can be interpreted as the probabilities $w_i$ for
rotations per unit time around each one of the four possible axis.
Frequencies smaller than 10$^{-6}$ ns$^{-1}$ correspond to less than
5 rotational jumps in 10 $\mu$s and therefore are negligible.

The tetrahedral constraint between the Cl
atoms imply that a $nn'$ transition can be achieved under two 
different single rotation events. Considering all rotation events
as independent processes, we can calculate $\nu_{nn'}$ as the
sum of the probability of the two possible events leading to a
$nn'$ transition. For example, $\nu_{12}=(w_3+w_4)/2$,
$\nu_{13}=(w_2+w_4)/2$, $\nu_{14}=(w_2+w_3)/2$ and so on. The
factor of $1/2$ is due to the assumption that both directions of
rotation are equally probably, which was tested to be statistically
true.

The next step is to solve the proposed master equation with the
four different initial conditions, which are $P^j_1(0)=(1,0,0,0)$,
$P^j_2(0)=(0,1,0,0)$, $P^j_3(0)=(0,0,1,0)$ and $P^j_4(0)=(0,0,0,1)$.
Let us denote the four corresponding solutions as $p^{j}_i(t)$,
with $i,j=1,\dots,4$. Then, the rotational auto correlation
functions can be expressed in terms of the solution of the master
equation as
\begin{equation}
\label{equation:master_correlations}
C_{j}(t)= \vec{b}_j \cdot \left[ \vec{b}_1 p^j_1(t) + \vec{b}_2 p^j_2(t) + \vec{b}_3 p^j_3(t) + \vec{b}_4 p^j_4(t) \right].
\end{equation}
The resulting solutions $C_j(t)$ are plotted on Figure
\ref{cgroups} using dashed lines. The agreement between the results
directly measured from the simulation trajectories and those
derived from the master equation is excellent, justifying the
assumptions made in the derivation of the analytical model.

\begin{figure}[t]
  \centering
  \includegraphics[width=0.4\textwidth]{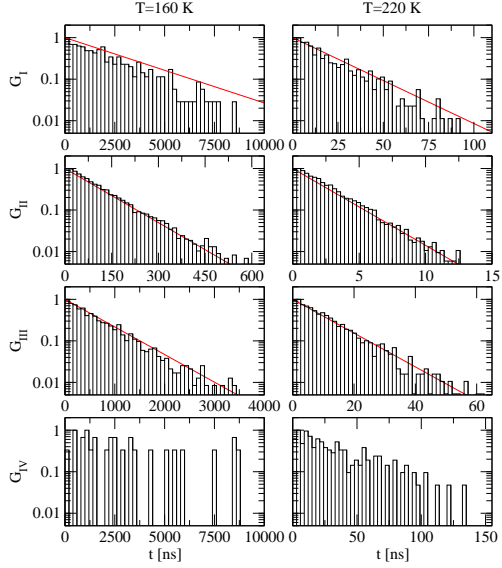}
  \caption{\label{hist_160_220} Distributions of waiting times for two
  temperatures and the corresponding Poisson curves (in red) for the fastest
  rotational axes of G$_{I}$, G$_{II}$ and G$_{III}$.}
\end{figure}

The statistics behind the rotational jumps can be evaluated by
studying the waiting time between successive jumps around the axes
$a_1,\dots,a_4$. On some of the axes the rotational events are very
rare and the length of our simulations do not allows to see a well
defined pattern for those specific slow axes. Nevertheless, for the
fastest axes the process reveals itself as following a Poisson
distribution. In Figure \ref{hist_160_220} we show the distribution
of waiting times for rotational events around the fastest axis of
rotation for each group of molecules. The histograms are directly
obtained from the simulated trajectories. The solid red lines in
the plots for the three fastest groups correspond to $e^{-\lambda t}$,
with the Poisson parameter $\lambda$ being the frequency of jumps
around $a_3$ axes, $w_3$ from Table \ref{table:w}.

\begin{figure}[t]
  \centering
  \includegraphics[width=0.4\textwidth]{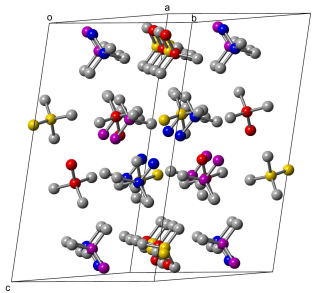}
  \caption{\label{parallel} Unit cell displaying the molecules of
groups G$_{I}$, G$_{II}$, G$_{III}$ and G$_{IV}$ in red,
blue, violet and yellow respectively.
The colored axis represent the fastest
direction of rotation for each molecule.}
\end{figure}

A similarity between the dynamical behavior of all the different
groups could be inferred by looking at the self correlation
functions $C_j(t)$ and distribution of waiting times shown in Figures
\ref{cgroups} and \ref{hist_160_220}, respectively. In particular,
G$_{II}$ and G$_{III}$ are the two groups
with the closest quantitative behavior. In order to explore whether
this kinetic resemblance has a correlation with the molecular
arrangement we analyzed the orientation of the fastest axis for
each one of the groups. Interestingly, the fastest axes of the
molecules of G$_{II}$ and G$_{III}$ are
parallel to each other in pairs that involve the first neighbors,
as shown in Figure \ref{parallel}. For groups G$_{I}$
and G$_{IV}$ there is a certain degree of spacial
correlation, but it does not include the eight pairs of molecules
of the unit cell.

NQR spectroscopy allows to measure the spin-lattice relaxation time
$T_1$. Since the low temperature crystalline structure of
CCl$_4$ has four nonequivalent molecules in the unit
cell, the NQR spectral lines correspond to the 16 Cl atoms of these
four molecules. Then it is possible to measure the relaxation
times $T_1$, for each line, as function of temperature. As nuclear magnetization
is proportional to the nuclear spin polarization, applying the slow
reorientational model for $T_1$ by Alexander and Tzalmona 
\cite{Alexander&Tzalmona} it is possible to find the reorientation
relaxation time as function of the jumps probabilities 
$\nu_{nm}$.

\begin{figure}[t]
\centering
\includegraphics[width=0.4\textwidth]{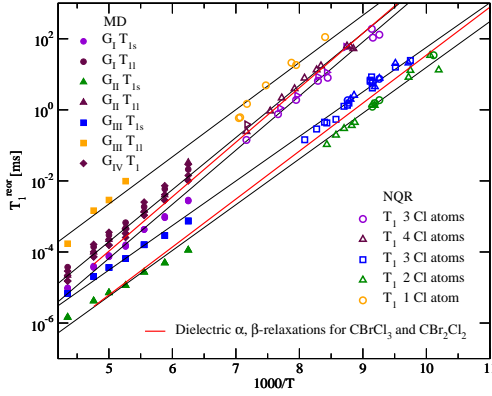}
\caption{\label{taus} Spin-lattice relaxation times obtained by
molecular dynamics simulations (full symbols). Results from NQR
studies of CCl$_4$ \cite{Zuriaga_JCP_2012} (empty symbols) are shown. Red lines represents the
behavior of $\tau_{\alpha}$ (upper) and $\tau_{\beta}$ (lower) of CBrCl$_3$
and CBr$_2$Cl$_2$ taken from Ref. \cite{Zuriaga_JCP_2012}.}
\end{figure}

Starting from the master equation for the nuclear polarization,
Zuriaga {\em et al} \cite{Zuriaga_JCP_2012} proposed a model
for tetrahedral molecules that allow us to find the chlorine $T_1$ relaxation
times as a function of reorientation frequencies $w_i$ around the four molecular
axes.
From the analytical model and the MD simulations, 
16 relaxation times are obtained at each temperature,
four for each group of molecules.
The results for $T_1$ vs. 1000/T are
represented in Figure \ref{taus} along with the experimental
results, which also include the cases of
CBrCl$_{3}$ and CBr$_{2}$Cl$_{2}$.

The existence of a preferential axis of rotation for three
groups (G$_{I}$, G$_{II}$ and G$_{III}$) allows the identification
of three short times  T$_{1s}$  of similar magnitude
(corresponding to the three Cl atoms out of the fastest axis)
and a long time T$_{1l}$ (for the Cl atom on the fastest axis).
The remaining group (G$_{IV}$) has no preferential axis of rotation
and therefore the four relaxation times are all similar.

The set of 16 relaxation times from the simulations are grouped
in increasing order in the following five sets:

\begin{enumerate}

\item The shortest time correspond to the T$_{1s}$ describing G$_{II}$ (green triangles).
\item The next time corresponds to T$_{1s}$ of G$_{III}$ (blue squares).
\item The third set is for T$_{1s}$ of G$_I$ (violet circles).
\item The fourth set includes T$_{1l}$ of G$_I$ (brown circles),
T$_{1l}$ of G$_{II}$ (brown triangles) and T$_1$ of G$_{IV}$ (brown diamonds).
\item The last set corresponds to T$_{1l}$ of G$_{III}$ (orange squares).
\end{enumerate}

The black straight lines in Figure \ref{taus}  are the result of a linear least-squares fit of the values $\log{T_1}$ vs. $1000/T$, including both, simulation and experimental results.

The are three main features that arise from Figure 5.
First, the simulations curves are in a good agreement with the experimental
data, prolonging the same Arrhenius curves from 100 to 230 K.
Second, the results for T$_1$ spread along two order of magnitude in time,
indicating the presence of fast and slow modes. Third, the overall behavior
resembles $\alpha$ and $\beta$ relaxation times of CBr$_2$Cl$_2$
(also shown in Figure \ref{taus}) and CBrCl$_3$,
showing that the dynamics of the non-glass former CCl$_4$ has strong
similarities with that of the other members of the isostructural
series CBr$_{n}$Cl$_{4-n}$.

\section{Conclusions}

We performed extensive molecular dynamics simulations of CCl$_4$ in
the monoclinic cell. By studying time reorientational auto-correlation
functions as a function of temperature we found distinct behaviors
for the four different C-Cl bonds in the molecule. As a consequence,
different re-orientational correlations times emerge for the four
groups of nonequivalent molecules in the system: two times are enough
to characterize three of the four groups and one time is associated
to the remaining group. 

The complete understanding of glassy systems, even in the case of 
the simpler rotational glasses, remains elusive in the sense that
no complete theoretical picture can be drawn yet. Nevertheless,
there is a general agreement on the existence of two main universal
features referred to $\alpha$ and $\beta$ relaxation. Rotational
glasses of the type CBr$_{n}$Cl$_{4-n}$, with
n=1, 2, 3 are glass formers displaying
fast and slow rotational modes as they approach to the glass
transition temperature. In this paper we show that
CCl$_4$ has essentially the same dynamical behavior,
as a function of temperature, to that of its isostructural glass
formers. Long molecular dynamics simulations trajectories, based on
a simple pair-wise additive force-field, yield to results that are
fully in line with those obtained with NQR experiments. The
simulations clearly show that there are preferential axes of
rotation, which are fixed with respect to the crystal orientation.
Two of the inequivalent group of molecules are significantly faster
than the other two, leading to a clear heterogeneity in the
dynamics of the system. Moreover, it is found that the orientation
of the two fast axes of rotations is the same, suggesting an
overall dynamics anisotropy correlated to the molecular
orientations.

Finally, our results on CCl$_4$ suggest that the observed heterogeneous
dynamic of molecules of type CXCl$_3$ with the same crystal structure
could be due to the molecular environment on the crystal
and not to the breaking of the tetrahedral symmetry of the molecule.
Then, it would be interesting to corroborate if there is any relation
between the fastest axis in the CCl$_4$ molecules and the C-Br axis in the 
CBrCl$_3$ crystals.
We are performing MD simulations to elucidate this open question.

\begin{acknowledgement}
 N.B.C., M.Z. and P. S acknowledge financial support of SECYTUNC and CONICET.
\noindent This work used computational resources from
CCAD Universidad Nacional de C\'ordoba
(http://ccad.unc.edu.ar/), in particular the Mendieta Cluster, which is part
of SNCAD MinCyT, Rep\'ublica Argentina.
\end{acknowledgement}

\begin{suppinfo}
A more detailed explanation of the method
for counting molecular jumps and C-C radial distribution functions at the lowest
and highest simulated temperatures along with experimental C-C distances are presented as
supplemental material. 
\end{suppinfo}

%
%

\section*{Dynamic Heterogeneity In The Monoclinic Phase Of CCl$_4$\\ \vspace{0.7cm} \underline{Supporting Information}}

\section*{Method for Detecting Rotational Jumps}

\noindent In order to detect the rotational jumps of the CCl$_4$ molecules
we perform a detailed analysis of the trajectories based on an algorithm
published by N. J. Carter and A. Cross; {\em Nature} {\bf 435} 308 (2005).
Let us consider the coordinates of each one of the Cl atoms, relative to their
bonded C atom, and denote them by $r_{n,\alpha,i}$. Here $\alpha$ represent
any of the Cartesian coordinates $x$, $y$ or $z$; the index $n$ denote
molecule and $i={1,2,3,4}$ refers to each one of the Cl atoms of molecule $n$.

\begin{equation}
{{f}}_{n,\alpha,i}(t)=\frac{\langle{r_{n,\alpha,i}}\rangle_- - \langle{r_{n,\alpha,i}}\rangle_+  }   {\sqrt{\frac{S_-}{\Delta t}+\frac{S_+}{\Delta t}}}
\end{equation}
where the angular brackets represent time average, the subindices $-$ and $+$
indicate the lapse of length $\Delta t$ before and after time $t$, respectively;
and $S_-$ and $S_+$ are the corresponding standard deviations.
Whenever there is a rotational jump the mean values and standard deviation are affected.
Consequently, the test function ${{f}}_{n,\alpha,i}(t)$ reflects those changes with
a spike. The spike could be upwards or downwards depending on the direction of
change of the average coordinate. \\

\noindent In order to have a single function per molecule, the twelve
individual ${{f}}_i(t)$ for an individual molecule can be combined in the
following way:
\begin{equation}
{\cal{F}}_n(t)= \sum_{i=1,2,3,4} \sum_{\alpha=x,y,z}  |{f_{n,\alpha,i}(t)}  |
\end{equation}

\begin{figure}
\centering
\includegraphics[width=0.65\textwidth]{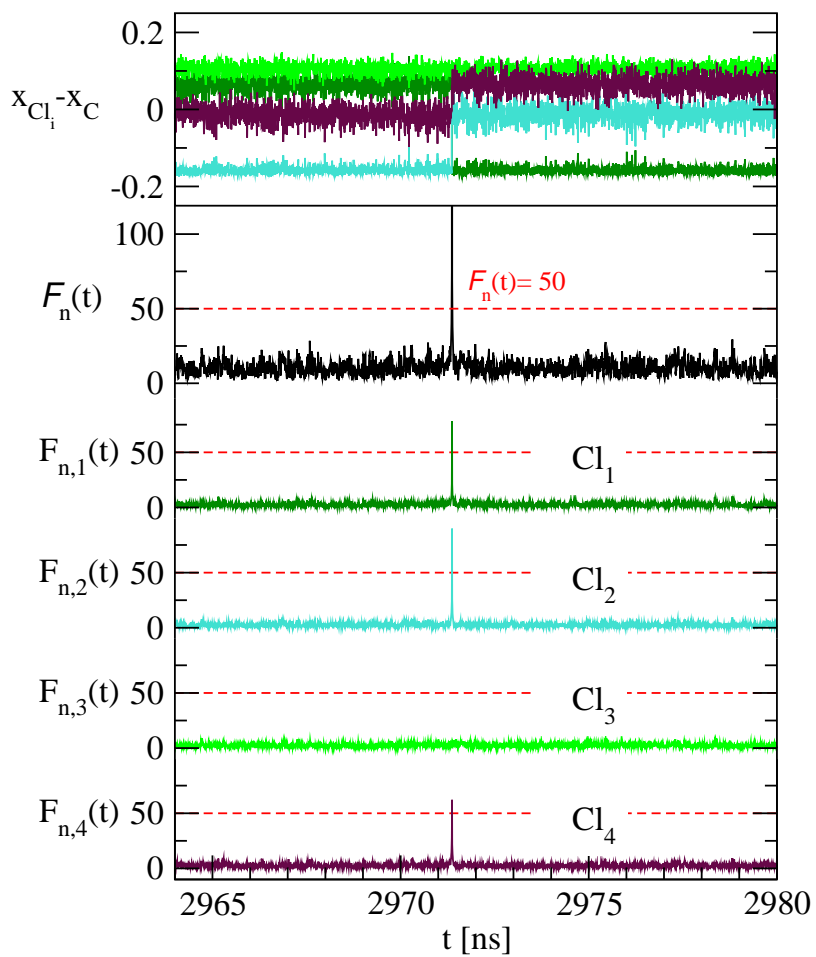}
\caption{ \label{test} Top: Variation of the relative x coordinates
of the four Chlorine atoms of molecule $n$ in green (Cl$_1$), cyan
(Cl$_2$), light green (Cl$_3$) and orange (Cl$_4$). The four
individual response functions are also shown in the corresponding
color, along with the final curve of the test function (in black).
The threshold value of 50 is shown in red.} 
\end{figure}

\noindent By careful inspection of the simulation trajectories and the
response of the test functions we selected a time lapse $\Delta t$= 50 ps.
In Figure \ref{test} we show an example of the variation of the relative
coordinates $x_{n,\alpha,i}(t)$ corresponding to the four Cl atoms of
molecule $n$, which undergoes a jump at $t=$2971.39 ns.
The figure also shows the corresponding four individual response functions of each
Cl atom:

\begin{equation}
{F}_{n,i}(t)= \sum_{\alpha=x,y,z}  |{f_{n,\alpha,i}(t)}  |
\end{equation}

along with the final curve of the test function ${\cal{F}}_n(t)$.
A threshold value of ${\cal{F}}_n(t)$=50 was used to determine the rotational jumps.

\section*{Radial Distribution Functions}

\begin{figure}
\centering
\includegraphics[width=0.65\textwidth]{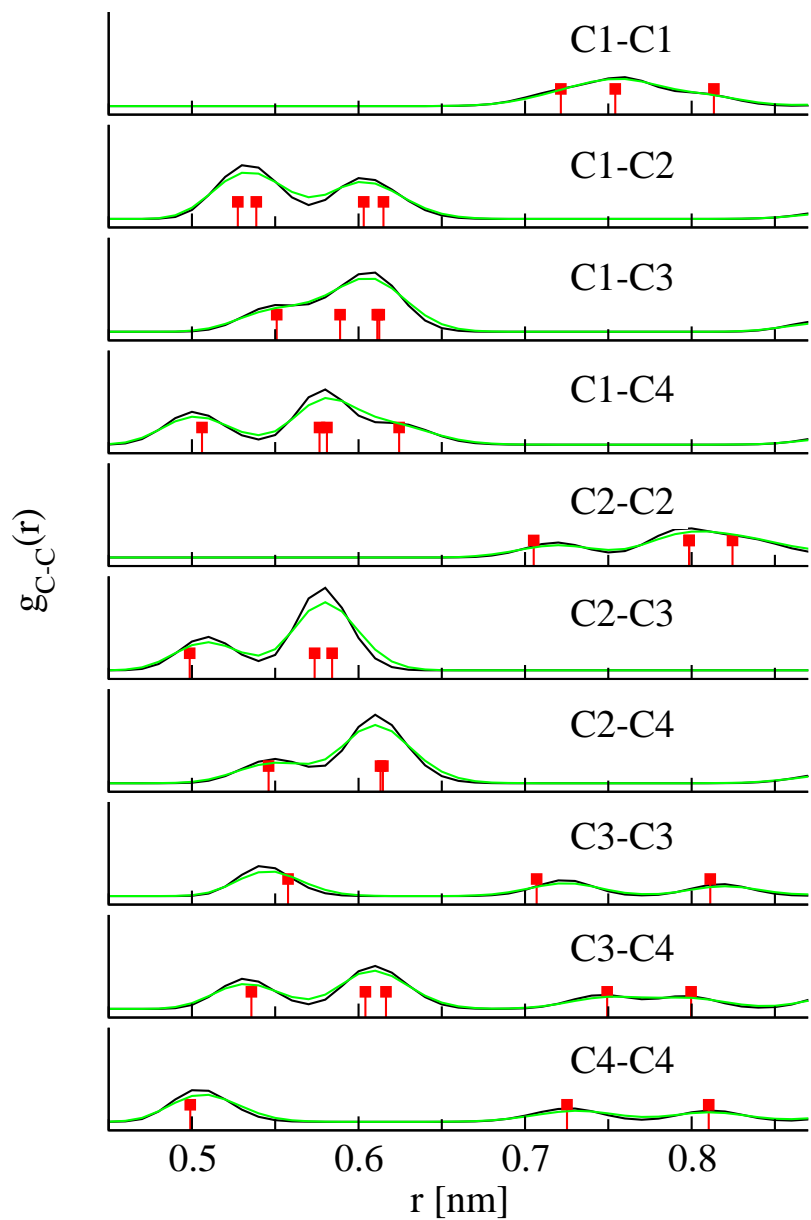}
\caption{Carbon-Carbon radial distribution function for all possible combination
of groups in the system at 160 K (black line) and 220 K (green line). On the graph
the experimental values are also shown in red.} \label{rdf}
\end{figure}

\noindent In order to analyze the crystalline structure of the simulation supercell,
we calculated the Carbon-Carbon radial distribution function for each possible
combination of Carbon atoms within the different 4 groups of molecules in the system,
at the all the simulation temperatures. We observed a good agreement of
the function's peaks with the experimental values for the Carbon-Carbon distances on
the monoclinic cell. On figure \ref{rdf} we show the results at the lowest and highest
simulation temperatures.

\end{document}